# Particle manipulation behind turbid medium based on intensity transmission matrix

**Kaige Liu,**[a,b,†] **Hengkang Zhang,**[c,†] **Shanshan Du,**[a,b] **Zeqi Liu,**[a,b] **Bin Zhang,**[d,*] **Xing Fu,**[a,b,*] **and Qiang Liu**[a,b,*]

[a]Key Laboratory of Photonics Control Technology (Tsinghua University), Ministry of Education, Beijing, China, 100084
[b]Tsinghua University, State Key Laboratory of Precision Measurement Technology and Instruments, Department of Precision Instrument, Beijing, China, 100084
[c]Beijing Institute of Control Engineering, Beijing, China, 100190
[d]Beijing Institute of Electronic System Engineering, Beijing, China, 100854

**Abstract**. Optical tweezers can manipulate tiny particles. However, the distortion caused by the scattering medium restricts the applications of optical tweezers. Wavefront shaping techniques including the transmission matrix (TM) method are powerful tools to achieve light focusing behind the scattering medium. In this paper, we propose a new kind of TM, named intensity transmission matrix (ITM). Only relying on the intensity distribution, we can calculate the ITM with only about 1/4 measurement time of the widely used four-phase method. Meanwhile, ITM method can avoid the energy loss in diffraction introduced by holographic modulation. Based on the ITM, we have implemented particle manipulation with a high degree of freedom on single and multiple particles. In addition, the manipulation range is enlarged over twenty times (compared with the memory effect) to 200 μm.

*Bin Zhang, E-mail: zhangbin1931@126.com
*Xing Fu, E-mail: fuxing@mail.tsinghua.edu.cn
*Qiang Liu, E-mail: qiangliu@mail.tsinghua.edu.cn
†These authors contributed equally.

## 1 Introduction

Optical tweezers use the interaction between light and matter to manipulate particles on the wavelength scale. The path of the photon is deflected as it passes through a particle, while the transfer of momentum forces the particle to move away from the direction of deflection. Thus, a Gaussian focused beam can create optical potential traps for binding or manipulating particles[1, 2]. Optical tweezers have achieved a wide range of applications in biomedicine[3-6], precision measurement[7-9], nanotechnology[10, 11], and many other fields[12-15]. Optical tweezers impose high demands on the stability of the system and the environment, and the light field distribution at the focal point has a great impact on the manipulation quality[16]. However, the prevalence of scattering media disrupts the distribution of the optical wavefront, making the focused beam a diffuse patch of light. Although scattered light fields can also be used to manipulate particles, the scattering effect causes the dispersion of energy and the irregularity of the spot shape, resulting in a less accurate and capable manipulation[17]. Therefore, achieving a high-quality optical focus behind turbid media will greatly expand the application range of optical tweezers.

In recent years, studies on the scattering effect have shown that light propagation in stable scattering media is a stationary process[18, 19]. Using wavefront shaping techniques, one can effectively compensate for the aberration induced by light scattering in turbid media to achieve focus or special light field distribution at the target location[20-23]. In 2010, Čižmár et al. used the stepwise sequence algorithm for phase optimization to achieve light focusing behind turbid medium and used this focus to implement particle manipulation[24]. In 2019, Peng Tong et al. used an improved genetic algorithm (GA) to achieve single-point and multi-point focusing and proceeded particle binding and manipulation[25]. They both relied on iterative optimization algorithms which set a specific target distribution of the light field and then optimized to approach it by iteratively changing the modulation mask based on the difference between the output light field and the targeted one. The memory effect of the scattering medium demonstrates that the output light will maintain the same motion when the input light is moved or angularly tilted in a small range[26, 27]. This feature can be used to move the light focus and thus manipulate the particles. However, the manipulation range of this method is usually small which is limited by the range of the memory effect. In addition, when dealing with multi-particle manipulation, each particle can only maintain the same movement as the input light, and cannot have its own individual trajectory, which greatly limits the freedom of multi-particle manipulation[25].

For particle manipulation, we usually wish to manipulate a large range of particles with a high degree of freedom and flexibility. We are aware that the transmission matrix (TM) method, another powerful tool of wavefront shaping, can achieve focus in large fields of view and multiple independent points[21, 23, 28-30]. It is an ideal tool for particle manipulation in scattering environments. However, the transmission matrix method requires the measurement of the response of a set of spatially complete orthogonal bases, which usually takes a lot of time and limits the optimization efficiency in practical applications[23]. This makes it difficult to apply this technique to dynamic media such as biological tissues. Therefore, a method that can quickly compute and obtain the transport matrix is still desired.

In this article, we propose a new kind of TM - intensity transmission matrix (ITM). The proposed ITM has two advantages over conventional TM. Firstly, we can measure the ITM of scattering media without the help of phase information. As a result, the measurement time of ITM reduced to about 1/4 compared to the four-phase method[21], greatly compressing the time consumption. On the other hand, it does not utilize the diffractive optical path, thus avoiding the loss of energy in diffraction and largely enhancing the intensity of focus[31]. In the remaining sections of this paper, we will first detail the theoretical framework of the ITM, including the theoretical prediction for the peak-to-background ratio (PBR) of the focus. Based on this theory, we implemented particle manipulation experiments, achieving both a large manipulation range and a high degree of freedom during multi-particle manipulation. We believe that this theory provides a new approach to make full use of the energy transported through the scattering medium with high speed. It will be significant to the applications of optical tweezers in a scattering environment.

## 2 Principles

The determination of the TM requires a set of spatially complete bases, and here we choose Hadamard bases. The two states of the DMD modulation cell correspond to 0 and 1 in amplitude modulation, so we replace all -1 elements in the Hadamard bases with 0 elements. The altered Hadamard bases are still a spatially complete matrix, and linear combinations of its column vectors can make up any modulation mask.

$$\varepsilon^{in} = H_{0-1} \cdot \alpha, \tag{1}$$

where $\varepsilon^{in}$ is an N-order vector recording the input modulation mask, $H_{0-1}$ is the altered Hadamard bases, $\alpha$ is an N-order vector.

To establish the relationship between the ITM and the altered Hadamard bases, we start with the conventional complex TM. In conventional TM theory, the measured output light field is in complex form, with[23]

$$\varepsilon^{tar} = E^{out} \cdot \alpha, \tag{2}$$

$$TM \cdot H_{0-1} = E^{out}, \tag{3}$$

$$TM \cdot H_{0-1} \cdot \alpha = E^{out} \cdot \alpha, \tag{4}$$

$$TM \cdot \varepsilon^{in} = \varepsilon^{tar}, \tag{5}$$

where $\varepsilon^{tar}$ is an M-order vector recording the output light field of the input $\varepsilon^{in}$, $E^{out}$ is an M×N matrix recording all the output light fields.

According to the TR, we can get the modulation mask corresponding to the target output[21]:

$$\varepsilon^{in} = TM^{\dagger} \cdot \varepsilon^{tar}, \tag{6}$$

where † denotes the transpose conjugate, and TM can be expressed as:

$$TM = E^{out} \cdot H_{0-1}^{-1} = \frac{2}{N} \cdot E^{out} \cdot \left( H - \frac{N}{2} M \right), \tag{7}$$

where $M$ is an N×N order matrix which has only one non-zero element in the first row and column.

Then we discuss the calculation of the ITM removing the phase information. The subtracted term on the right side of equation (7) only affects the magnitude of the first column of TM, without changing its sign. And it has no effect on the elements of the other columns. Therefore, this term is not necessary for determining the effect of a module in DMD when calculating the optimization mask and can be ignored.

The mask corresponding to focus at the $m$-th output mode is obtained from the $m$-th row in TM. And since the output light field $E^{out}$ is complex-valued, the target output $\varepsilon^{tar}$ obtained from the linear combination of $\alpha$ is consistent with the coherent superposition in the optical transmission process, and the calculated input mask is also the theoretically optimal one[23]. However, if no phase measurements are made, the TM is calculated by relying only on the intensity distribution. That is to say, $E^{out}$ is real-valued. And it is important to note that the time inversion still holds, since it is equivalent to taking a mode on both sides of the equation, and the equation still holds when both are real numbers.

To verify the effectiveness of ITM in wavefront shaping, we first consider the distribution of elements in the ITM:

$$ITM = \frac{2}{N} \cdot \left| E^{out} \right| \cdot H, \tag{8}$$

where $H$ is the conventional Hadamard bases.

If our target focus is placed at the $m$-th output mode, we still extract the $m$-th row of the ITM:

$$\varepsilon^{in} = \frac{2}{N} \cdot \left( |e_{m1}|, |e_{m2}|, \cdots, |e_{mN}| \right) \cdot H, \tag{9}$$

where $|e_{mj}|$ is the amplitude of the $m$-th output mode with the $j$-th column vector in Hadamard bases entering. Then we decide which input mode should be open based on its influence on the intensity at target output mode:

$$(t_{m1}, t_{m2}, \cdots, t_{mN}) = \frac{2}{N} \cdot (|e_{m1}|, |e_{m2}|, \cdots, |e_{mN}|) \cdot H, \tag{10}$$

$$t_{mj} = \frac{2}{N} \cdot (|e_{m1}|, |e_{m2}|, \cdots, |e_{mN}|) \cdot h_j = \frac{2}{N} \sum_{i=1}^{N} h_{ij} |e_{mi}| = \frac{2}{N} \sum_{h_{ij}=1} |e_{mi}| - \frac{2}{N} \sum_{h_{ij}=-1} |e_{mi}|, \tag{11}$$

where $t_{mj}$ is the element of the $m$-th row and $j$-th column in the ITM, $h_j$ is the $j$-th column vector in Hadamard bases. Here $h_{ij}$ takes 1 and -1 corresponding to 1 and 0 in the altered Hadamard bases, respectively. For example, if $h_{kj} = -1$ when $i = k$, then it corresponds to the case where the $j$-th input module is off when the $i$-th Hadamard vector is input. At this point, the light field at the $j$-th input module is not involved in the intensity superposition at the target output mode. Therefore, $|e_{mk}|$ is not affected by the light field at the $j$-th input module. That is, all the $|e_{mj}|$ terms corresponding to the value of $i$ that match $h_{ij} = -1$ are obtained with the $j$-th input module turned off. On the contrary, the ones corresponding to $h_{ij} = 1$ are obtained while the $j$-th input module is turned on. For Hadamard bases, open modules and close modules are split in half disregarding the first input mode. Therefore, the equation (11) is converted into the difference of the mean value of the amplitude at the target output mode when the $j$-th input module is turned on and off. Since N is much larger than 1, the input modules other than the $j$-th input module can be regarded as randomly turned on and off in both cases. Therefore, the above equation can be translated to:

$$t_{mj} = \langle |a + a_1 + a_j + i(b + b_1 + b_j)| \rangle - \langle |a + a_1 + i(b + b_1)| \rangle, \tag{12}$$

where $t = a + ib$ is one of the TM elements and $t_{m1} + t_{mj} = a_1 + a_j + i(b_1 + b_j)$ (the first module is always open), $\langle \ \rangle$ donates the statistical average.

When $t_{mj} > 0$, the opening of the $j$-th input module causes the mean value of the intensity at the $m$-th output mode to be boosted. So, the $j$-th input module has a positive effect on the focus and should be opened. Based on the sign of $t_{mj}$, we can qualitatively estimate the effect of all the input modules on the target output mode and thus obtain the input mask to achieve the focusing.

Finally, in order to evaluate ITM's ability to focusing, we can use the statistical properties of the scattering media to calculate the theoretical value of PBR[22]. Under the condition of $t_{mj} > 0$, equation (12) is equivalent to:

$$\langle |a+a_1+a_j+i(b+b_1+b_j)| \rangle > \langle |a+a_1+i(b+b_1)| \rangle, \tag{13}$$

$$\langle |a+a_1+a_j+i(b+b_1+b_j)|^2 \rangle > \langle |a+a_1+i(b+b_1)|^2 \rangle, \tag{14}$$

According to the first-order statistical properties of the scattered light field, we can obtain the joint probability density function of the real and imaginary parts of each point in the scattered light field[32]:

$$p_{r,i}(r,i) = \frac{1}{2\pi\sigma^2} \exp\left(-\frac{r^2+i^2}{2\sigma^2}\right), \tag{15}$$

where $r$ and $i$ are respectively the real and imaginary parts of the output light field, and $\sigma$ is the variance of the Gaussian distribution. Then we can calculate the values on both sides of the equation (14):

$$\begin{aligned}
&\langle |a+a_1+i(b+b_1)|^2 \rangle \\
&= \iint_\infty \frac{(a+a_1)^2+(b+b_1)^2}{2\pi\sigma^2} \exp\left(-\frac{a^2+b^2}{2\sigma^2}\right) dadb \\
&= \iint_\infty \frac{a^2+b^2}{2\pi\sigma^2} \exp\left(-\frac{a^2+b^2}{2\sigma^2}\right) dadb + \iint_\infty \frac{a_1^2+b_1^2}{2\pi\sigma^2} \exp\left(-\frac{a^2+b^2}{2\sigma^2}\right) dadb \\
&+ \iint_\infty \frac{2aa_1+2bb_1}{2\pi\sigma^2} \exp\left(-\frac{a^2+b^2}{2\sigma^2}\right) dadb \\
&= 2\sigma^2 + a_1^2 + b_1^2
\end{aligned} \tag{16}$$

$$\langle |a+a_1+a_j+i(b+b_1+b_j)|^2 \rangle = 2\sigma^2 + (a_1+a_j)^2 + (b_1+b_j)^2. \tag{17}$$

If the module has a positive effect, we can get

$$(a_1+a_j)^2 + (b_1+b_j)^2 > a_1^2 + b_1^2. \tag{18}$$

Here $t_{m1}$ is taken as the statistical average:

$$|t_{m1}| = \sqrt{\frac{\pi}{2}}\sigma, \varphi_{m1} = 0, \tag{19}$$

$$a_1 = \sqrt{\frac{\pi}{2}}\sigma, b_1 = 0, \tag{20}$$

where $|t_{m1}|$ and $\varphi_{m1}$ are the amplitude and phase of the $t_{m1}$, respectively. Therefore, the value of $t_{mj}$ needs to satisfy:

$$\left(a_j + \sqrt{\frac{\pi}{2}}\sigma\right)^2 + b_j^2 > \frac{\pi}{2}\sigma^2. \tag{21}$$

From the probability density function (equation (15)), the statistical characteristics of $t_{mj}$ can be calculated:

$$p_{open} = \iint_{\left(x+\frac{\sqrt{\pi}}{2}\sigma\right)^2+y^2>\frac{\pi}{2}\sigma^2} \frac{1}{2\pi\sigma^2}\exp\left(-\frac{x^2+y^2}{2\sigma^2}\right)dxdy \approx 0.5809, \tag{22}$$

$$\langle|t_{mj}|\rangle = \frac{1}{p_{open}}\iint_{\left(x+\frac{\sqrt{\pi}}{2}\sigma\right)^2+y^2>\frac{\pi}{2}\sigma^2} \frac{\sqrt{x^2+y^2}}{2\pi\sigma^2}\exp\left(-\frac{x^2+y^2}{2\sigma^2}\right)dxdy \approx 1.5252\sigma, \tag{23}$$

$$\langle|t_{mj}|^2\rangle = \frac{1}{p_{open}}\iint_{\left(x+\frac{\sqrt{\pi}}{2}\sigma\right)^2+y^2>\frac{\pi}{2}\sigma^2} \frac{x^2+y^2}{2\pi\sigma^2}\exp\left(-\frac{x^2+y^2}{2\sigma^2}\right)dxdy \approx 2.7297\sigma^2, \tag{24}$$

$$\langle\exp(i\varphi_{mj})\rangle = \frac{1}{p_{open}}\iint_{\left(x+\frac{\sqrt{\pi}}{2}\sigma\right)^2+y^2>\frac{\pi}{2}\sigma^2} \frac{x+iy}{2\pi\sigma^2\sqrt{x^2+y^2}}\exp\left(-\frac{x^2+y^2}{2\sigma^2}\right)dxdy \approx 0.3828, \tag{25}$$

where $p_{open}$ is the possibility for any module to open, $|t_{mj}|$ and $\varphi_{mj}$ are respectively the amplitude and phase of the $t_{mj}$. Thus, the focus intensity at the $m$-th output mode can be derived from the superposition of the complex amplitudes of the open modules as follows:

$$I^{foc} = \left|\sum_{n=1}^{N_{open}} At_{mn}\right|^2 = \sum_{n=1}^{N_{open}} A^2 |t_{mn}|^2 + \sum_{n \neq h}^{N_{open}} A^2 t_{mn}^* t_{mh}, \tag{26}$$

where $A$ is the normalization factor. The statistical mean of the focus points is

$$\langle I^{foc} \rangle = A^2 \sum_{j=1}^{N_{open}} \langle |t_{mj}|^2 \rangle + A^2 \sum_{j \neq k}^{N_{open}} \langle |t_{mj}| \rangle \langle |t_{mk}| \rangle \langle \exp(i\varphi_{mj}) \rangle \langle \exp(i\varphi_{mk}) \rangle$$

$$= A^2 \times 0.5809 N \left[ 2.7297\sigma^2 + (0.5809 N - 1) \times 1.5252^2 \sigma^2 \times 0.3828^2 \right] \tag{27}$$

The average intensity of the background field is

$$\langle I^{bg} \rangle = A^2 \sum_{j=1}^{N_{open}} \langle |t_{mj}|^2 \rangle = A^2 \times 0.5809 N \times 2.7297\sigma^2, \tag{28}$$

and the focus PBR is

$$PBR = \frac{\langle I^{foc} \rangle}{\langle I^{bg} \rangle}$$
$$= \frac{A^2 \times 0.5809 N \left[ 2.7297\sigma^2 + (0.5809 N - 1) \times 1.5252^2 \sigma^2 \times 0.3828^2 \right]}{A^2 \times 0.5809 N \times 2.7297\sigma^2} \tag{29}$$
$$= 1 + 0.1249(0.5809 N - 1)$$
$$\approx 0.0726 N$$

The theoretical PBR for binary modulation using a complex form transmission matrix is[22]

$$PBR = 1 + \frac{1}{\pi}\left(\frac{N}{2} - 1\right) \approx 0.1592 N. \tag{30}$$

The theoretical PBR of the ITM is about 1/2 of the conventional one.

## 3 Experiment

### 3.1 Experimental setup

In our experiment, a 532 nm laser (home-made, solid-state, continuous-wave) with a maximum output power of 6 W was employed as the light source. It was first expanded by a beam expander and then went through the combination of a half-wave plate and a polarizer to modulate the polarization state. The modulator was a high-speed DMD (ViALUX, V7001, resolution: 1024 × 768, pixel size: 13.7 μm × 13.7 μm), which selected parts of the beam to reflect into the subsequent system. A 4-f telescope system (f1 = 600 mm, f2 = 300 mm) and a 50:50 beam splitter (BS) were employed to image the modulation mask on the aperture of the objective (Obj1, Nikon, TU Plan ELWD 50×, NA=0.6). Then the modulated beam was focused onto the scattering medium by the objective. The scattering medium was a ground glass scatterer (Edmund, 120 grits), through which the light was scattered. The scattered light then passed through the particle sample (5 μm diameter, polystyrene) in water accommodated by a slide and coverslip. Another objective (Obj2, Olympus, MPLAN-N 20×, NA = 0.4) and a tube lens (f3 = 180 mm) were used to image the transmitted light to a complementary metal oxide semiconductor (CMOS) camera (PCO, edge 4.2 bi, pixel size: 6.5 μm × 6.5 μm), and used as the microscopic objective for the observation of particle sample motion. The LED was utilized to provide illumination while manipulating the particle. When observing the particle movement, the 532 nm light was blocked by a 532 nm centered band-blocking filter to improve the observation contrast. The control program for the camera and DMD was written in C++ (Visual studio 2019 community).

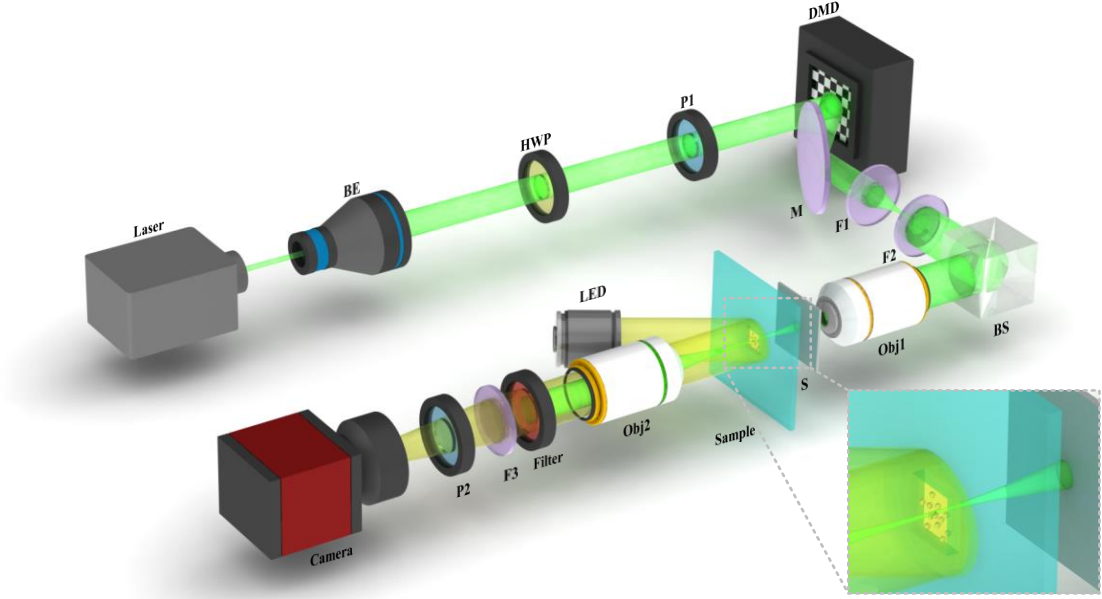

**Fig. 1** Experimental setup. F1, F2, F3: lens, HWP: half-wave plate, P1, P2: polarizer, BE: beam expander, M: mirror, BS: beam splitter, S: scattering medium, Obj: objective lens.

*3.2    Intensity transmission matrix*

We first experimentally verified the ability of the ITM in light focusing behind the scattering medium. The ITM was measured using the altered Hadamard bases of order 4096, indicating that the number of modulation cells was N = 4096. Using TR, an optimized mask for focusing at the target output mode could be calculated. Then we loaded the mask on the DMD, and the focused spot was observed on the image plane with a PBR of 215.66. And the theoretically predicted value could be calculated according to the previously-inferred theoretical value of PBR by equation (29), i.e. 0.0726×4096+0.8751=298.24. In addition, this method can also achieve multi-point focusing when the target output light field increased to several points. The results were shown in Fig. 2(c). The corresponding PBR at three focal points were 71.27, 70.13, and 75.59, respectively.

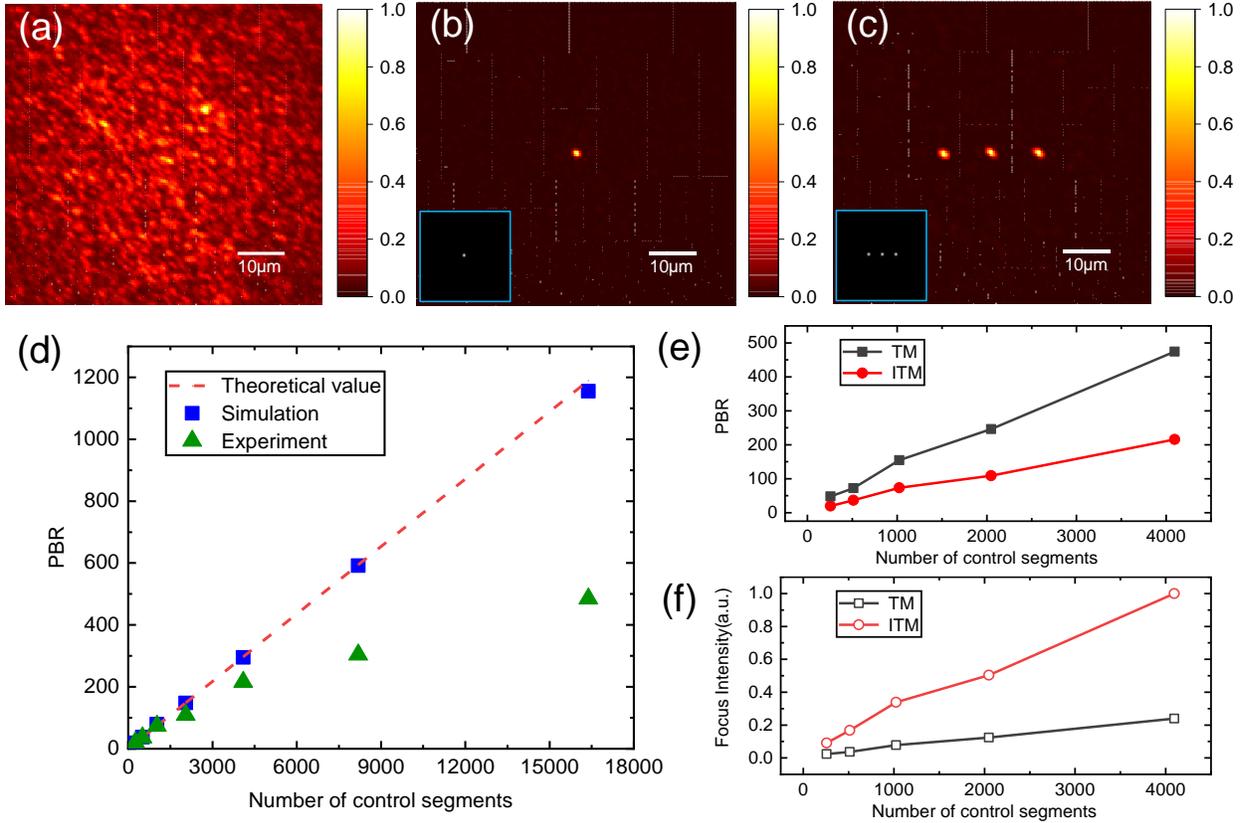

**Fig. 2** Focusing result with ITM, (a): speckle pattern before optimization, (b): experimental result and target output light field of single focus, (c): experimental result and target output light field of multiple foci, (d): PBR with different control segments number in theory, simulation, and experiment, (e): comparison of TM and ITM in terms of PBR, (f): comparison of TM and ITM in terms of focus intensity.

When we changed the control segments number and performed focusing experiments using different orders of Hadamard bases, we obtained the curve of PBR, as shown in Fig. 2 (d). We also used the angular spectrum method[33] to simulate the process of light focusing and obtained the simulation result. When the control segments number was small, the PBR agreed well with the theoretical prediction. While when the control segments number was large, the PBR of the focus was lower than the theoretical value. This error was attributed to the speckle decorrelation[34] and the measurement noise in the experiment[22].

On the other hand, the focal intensity is also an important factor to be considered to achieve particle manipulation behind scattering media. We compared the PBR and peak intensity of the focal point of ITM and TM in our experiments and the results were shown in Fig. 2(e) and Fig. 2(f), respectively. Although the theoretical value of the PBR of ITM is lower than that of TM by about 1/2, the energy loss in transmission was greatly reduced by the ITM compared with the conventional TM using holographic modulation due to its diffraction-free optical path. The combined effect resulted in a several-fold increase in the focal intensity of ITM compared to TM. This improvement reached 4.2 times at N=4096. In addition to this intensity improvement, ITM compresses the measurement time to about 1/4 times because it does not require phase measurement.

*3.3    Particle manipulation*

The ITM method allows for rapid acquisition of single-point and multi-point focus close to the diffraction limit. The large light intensity gradient can be used to capture particles undergoing Brownian motion in liquid. The particles can be captured by the focus as they move randomly near the focus and will be stably bound to the focus[1, 2]. As the focus moving, the particle will be dragged to the center of the focus continuously, thus moving in the same path of the focus.

After the transmission matrix was measured, we calculated the optimized masks for focusing at different output modes and then loaded these masks sequentially onto the DMD to achieve arbitrary trajectory movement of the focal point on the image plane. With this method, we could move the focus in a large range. Without the limitation of memory effect, the range of the manipulation could be enlarged to 200 μm. We pulled the particle in a straight line for about 200 μm (see Video 1), as shown in Fig. 3. And with a larger imaging area and more output modes, the range of manipulation could be further expanded.

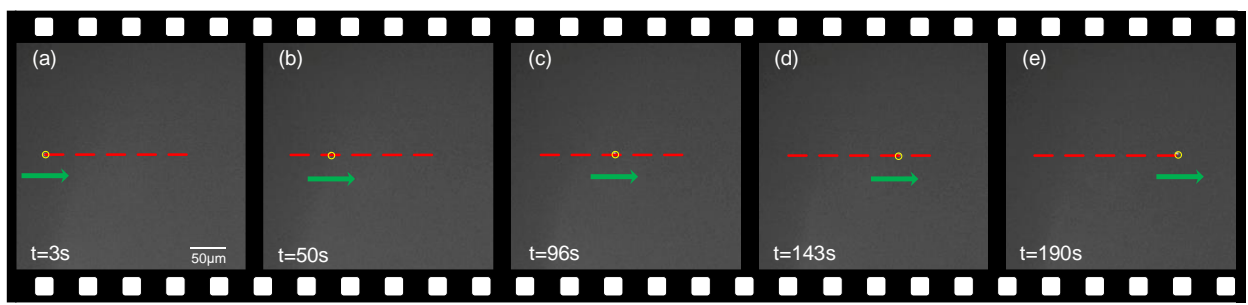

**Fig. 3** Particle manipulation with single particle along a straight line.

Figs. 4(a)-4(e) showed particle manipulation with special trajectories (see Video 2). In addition, using the results of multi-focusing, we could manipulate multiple particles at the same time. When calculating the optimization mask, the positions of the multiple targets were calculated independently, allowing the multi-focus to move independently with their own trajectories. When performing multi-particle manipulation, each particle could operate independently and would not be affected by the other particles. Figs. 4(f)-4(j) showed the result of manipulating two particles moving around a circle (see Video 3).

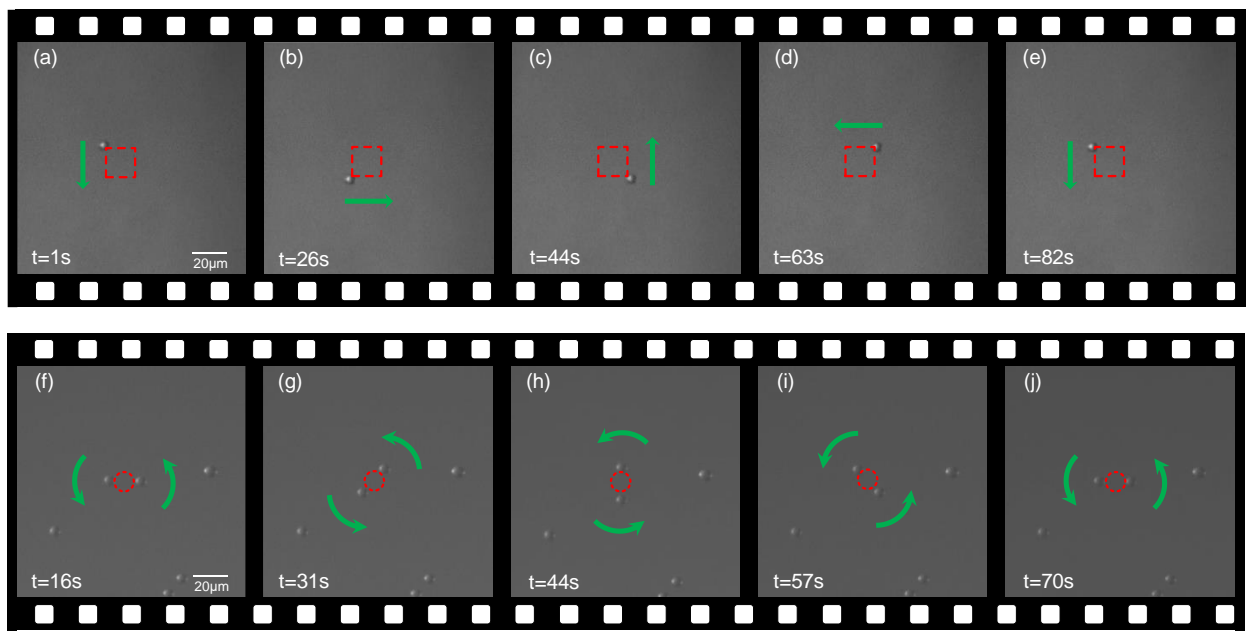

**Fig. 4** Particle manipulation, (a)-(e): single focus with the square path, (f)-(j): double foci with the circle path.

## 4   Discussion

During the process of particle manipulation, fluctuations of the focus intensity cause changes in the constraint stiffness. When the constraint stiffness decreases, the particles will have a higher probability of escaping the bindings[35]. The focus intensity of the iterative algorithm depends on the distance moved relative to the original focus. As the distance increases, the intensity decreases smoothly[25]. In contrast, the focus achieved by ITM does not depend on the distance moved but is limited by the inhomogeneity of the focus intensity at different points.

To evaluate the stability of the particle manipulation, we first measured the focus intensity at different output modes by taking points at intervals in a large imaging area, as shown in Fig. 5(a). In the central part of the imaging area (the blue square in Fig. 5(a)), we take points continuously and the distribution of PBR was shown in Fig. 5(b). Both interval and continuous points were recorded with 64×64 pixels. We calculated the mean and variance of the PBR in circular regions with different radius in pixels, as shown in Fig. 5(c). As the radius increased, the mean value within the region showed a decreasing trend. This was because the beam was first focused by the objective lens and then scattered by the scattering medium. The concentration of energy still tended to be in the form of a Gaussian distribution. Therefore, the closer to the edge part of the image plane, the less energy could be optimally converged, and the lower the peak of the focus obtained. The variance in the region stayed at a low level, almost constant with increasing radius. This was due to the inhomogeneity of the focus intensity distribution caused by the inhomogeneity of the incident light field and the errors introduced by environmental perturbations. The variance of continuously recording points was lower overall compared to recording points with intervals. Although the variance of both was close at a large pixel radius, continuously recording points could get very low variance at a small pixel radius. Therefore, the continuous output modes could reduce the instability in the particle manipulation process. In addition, the continuous output modes had a smaller spatial distance between modes and were less likely to lose particles during focus shifting.

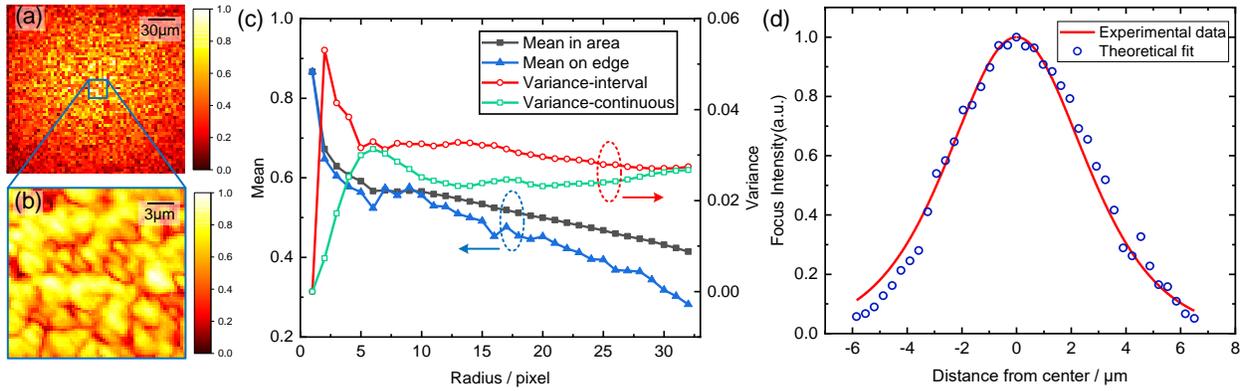

**Fig. 5** Move focus with ITM and memory effect, (a) (b): uniformity of PBR focused by ITM (64×64 order) at different output modes with recording one point in ten points and recording points continuously, respectively, (c): mean and variance of the focus PBR with ITM, (d): move focus with memory effect.

The intensity at the edge of the circular region determines the range of particle manipulation. It shows a gradual decrease along the radial direction and drops to half at the radius of about 68 μm. And at the radius of 104 μm, the mean PBR decreases to 32.5%. The farther the scattering medium is placed from the focal point of the focusing objective, the larger this width becomes. But it also leads to a decrease in the focal energy. Therefore, when performing particle manipulation, we need to choose appropriate system parameters to obtain the best manipulation performance. In addition, we tested the range of movement of the memory effect under our experimental conditions. By shifting the position of the beam entering the scattering medium within a small range, we observed a following shift of the focal point. The further away from the original position, the lower the intensity was, as shown in Fig. 5(d). Based on the formula for the memory effect[26, 27]:

$$I(\theta, L) = k_0 \theta L / \sinh(k_0 \theta L), \tag{31}$$

where $I(\theta, L)$ is the focus intensity, $k_0$ is the wave factor, $\theta$ is the angle of deviation from the initial position, $L$ is the effective thickness, we fitted the theoretical curve. Its full width at half maxima (FWHM) was 6.2 μm, and the intensity decreased to less than 1/3 of the peak value outside the 7.5 μm range. Therefore, compared to the memory effect, the manipulation range could be extended by more than twenty times using ITM.

## 5 Summary

In this paper, we have proposed a new kind of transmission matrix, named intensity transmission matrix. It can be calculated without phase measurement, thus avoiding the energy loss in holographic modulation. It also compresses the measurement time to about 1/4 compared to the conventional four-phase method. Based on the ITM, we had achieved single-point and multi-point focusing. Then we applied the ITM to particle manipulation taking advantage of its large range and a high degree of freedom in manipulation. The manipulation range was enlarged over twenty times to about 200 μm compared with the memory effect. Furthermore, multiple particles could be manipulated at the same time with their own trajectories.

*Disclosures*

The authors declare no conflicts of interest.

*Acknowledgments*

This work was supported by the National Natural Science Foundation of China (NSFC) (61905128, 61875100).

**Kaige Liu** is a graduate student at Tsinghua University. He received his BS degree in measurement and control technology and instrumentation from Tsinghua University in 2020. His current research interests include wavefront shaping, transmission matrix, structured light, neural networks and optical manipulation.

**Hengkang Zhang** received his BS degree in electronic science and technology from Harbin Institute of Technology in 2015, and his PhD degree in optical engineering from Tsinghua University in 2021. His current research interests include wavefront shaping, transmission matrix, structured light and optical manipulation.

**Bin Zhang** received his BS degree in electronic science and technology from Beijing University of Chemical Technology in 2013, and his PhD degree in control science and engineering from Beijing University of Chemical Technology in 2018. From 2019 to 2021, he was a postdoctoral fellow at Tsinghua University. His current research interests include wavefront shaping, fluorescence imaging, and opto-acoustic imaging.

**Qiang Liu** is a professor at Tsinghua University and the director of the Key Laboratory of Photonics Control Technology. He received his PhD degree in physical electronics from Harbin Institute of Technology in 2001. And he was selected as Changjiang Scholar Distinguished Professor in 2017. He has been working on the research of high-power solid-state laser and nonlinear frequency-conversion technology, now he has authored more than 100 journal papers.

His research interests also include spatial modulation of light fields, biomedical imaging, and optical manipulation.

Biographies and photographs for the other authors are not available.

**Caption List**

**Fig. 1** Experimental setup. F1, F2, F3: lens, HWP: half-wave plate, P1, P2: polarizer, BE: beam expander, M: mirror, BS: beam splitter, S: scattering medium, Obj: objective lens.

**Fig. 2** Focusing result with ITM, (a): speckle pattern before optimization, (b): experimental result and target output light field of single focus, (c): experimental result and target output light field of multiple foci, (d): PBR with different control segments number in theory, simulation, and experiment, (e): comparison of TM and ITM in terms of focus intensity.

**Fig. 3** Particle manipulation with single particle along a straight line.

**Fig. 4** Particle manipulation, (a)-(e): single focus with the square path, (f)-(j): double foci with the circle path.

**Fig. 5** Move focus with ITM and memory effect, (a) (b): uniformity of PBR focused by ITM (64×64 order) at different output modes with recording one point in ten points and recording points continuously, respectively, (c): mean and variance of the focus PBR with ITM, (d): move focus with memory effect.